\documentclass[%
 reprint,
 amsmath,amssymb,
 aps,
pra,
longbibliography,
]{revtex4-1}

\usepackage{natbib}
\usepackage{graphicx}
\usepackage{dcolumn}
\usepackage{bm}
\usepackage[colorlinks=true ,citecolor=blue]{hyperref}

\begin{document}


\author{S\'alvio Jacob Bereta}
 \email{salvio.bereta@usp.br}
\affiliation{Instituto de F\'\i sica de S\~{a}o Carlos, Universidade de S\~{a}o Paulo, Brazil }

\author{M\^{o}nica A. Caracanhas}
\affiliation{Instituto de F\'\i sica de S\~{a}o Carlos, Universidade de S\~{a}o Paulo, Brazil}%

\author {Alexander L. Fetter}
\email{fetter@stanford.edu}
\affiliation {Departments of Physics and Applied Physics, Stanford University, Stanford, CA 94305-4045, USA}

\title{Superfluid vortex dynamics on a  spherical  film}





\date{\today}

\begin{abstract}

Motivated by ongoing experimental efforts to make closed Bose-Einstein condensate (BEC) shells in microgravity environments, this work studies the energy and dynamics of singly quantized vortices on a thin spherical superfluid shell, where the overall vortex charge must vanish (as on any compact surface).  For each vortex, stereographic projection yields the corresponding complex potential on the tangent plane.  The resulting stream function then provides both the total energy and the dynamics of a system of overall neutral vortices on a spherical film.  Although a single vortex dipole follows a simple dynamical orbit,  four vortices can present a variety of situations.  We study a few symmetric initial configurations and then focus on the special case of two small vortex dipoles.

\end{abstract}

\date{\today}


\pacs{Valid PACS appear here}
\maketitle


\section{Introduction}

The remarkable creation (1995) of  Bose-Einstein condensates (BECs) in ultracold dilute atomic gases has renewed interest in bosonic superfluids (see, for example,  Refs~\cite{Pethick,Pitaevskii2016}).  These highly  flexible many-body coherent systems now appear in many different forms, for example, three-dimensional bulk condensates ranging from flat pancakes to elongated cigars,  and various optical lattices.  They also provide a basis for tests of fundamental quantum matter Ref.~\cite{Bloch}. 

Recent   proposals to create thin spherical traps in space (leading to ``bubble" BECs) Refs.~\cite{NASA,LundbladandCarollo} along with the related works in Refs.~\cite{13,7} have stimulated research on thin-film superfluid dynamics on curved surfaces Ref.~\cite{Nils-Eric,Pietro}.   Here we study thin spherical superfluids that we model as ideal classical fluid films with quantized vortices.  In addition, the compact spherical surface requires that the system have zero net vorticity.  

Most investigations have emphasized the energy of quantized point vortices on general curved surfaces (see, for example, Ref.~\cite{Turner}), usually based on the phase $S(\bm r)$ of the superfluid  condensate wave function.    
In contrast, we  focus  here on   the dynamics of these vortices and specialize to the surface of a sphere.  

To this end, we rely on the stream function $\chi(\Omega)$ at a point with spherical polar angular coordinates $\Omega = (\theta,\phi)$.  In addition, $\chi$   also depends parametrically on the location of the various point vortices at $\Omega_j$.   Not only does $\chi$ determine directly the motion of the set of vortices, but it also immediately gives the interaction energy of the same vortices with no additional analysis.  A combination of these results generalizes the familiar Hamiltonian  formulation of vortex dynamics on a plane  to the curved surface of a sphere, with the angular coordinates of a vortex  $\Omega_j = (\theta_j,\phi_j)$ as canonical Hamiltonian variables (see, for example, Sec.~157 of Ref.~\cite{Lamb}).

In Sec.~II, we summarize vortex dynamics in a plane, where we review the complementary roles of the condensate phase $S$ (which is effectively the velocity potential) and of the stream function $\chi$ in determining the local flow velocity $\bm v(\bm r)$.  For the  surface of a sphere (Sec.~III), we use a stereographic projection to determine the corresponding stream function $\chi(\Omega)$ and hence the vortex dynamics on a sphere. As an example, we study  the dynamics of a single  vortex dipole.  Section IV obtains the energy of a set of vortices on a sphere in terms of the same stream function, which then
provides a Hamiltonian formalism for the vortex dynamics.   Section~V studies  the dynamics of four vortices in a few simple symmetric configurations.   The energy and dynamics of two small  vortex dipoles are considered in Sec.~VI, and Sec.~VII provides a summary and conclusions.

\section{Superfluid vortex dynamics on a plane}

Typical superfluids like He-II and atomic BECs have scalar complex order parameters $\Psi = |\Psi| e^{iS}$.  The  associated superfluid velocity is 
\begin{equation}\label{nablaS}
\bm v = (\hbar/M) \bm \nabla S,
\end{equation}
 where $2\pi \hbar = h$ is Planck's constant and $M$ is the  relevant atomic mass.  Apart from an overall factor, the phase $S(\bm r)$ is the velocity potential.  Equation~(\ref{nablaS}) implies that the superfluid velocity field is irrotational with  $\bm \nabla \times \bm v = 0$ except at singular points that represent  quantized vortices.  Each vortex has quantized  circulation 
\begin{equation}
 \oint_{\cal C} d\bm l\cdot \bm v =\frac{\hbar}{M}\oint_{\cal C} d\bm l \cdot \bm \nabla S =\frac{2\pi \hbar}{M}  \nu,\end{equation}
where $\nu$ is the  winding number of the phase around the closed contour $\cal C$.   The fundamental unit of circulation is $ 2\pi\hbar/M$, which has the dimension of length squared over time, like a diffusivity or a kinematic viscosity.

In many cases, the fluid is also incompressible with $\bm\nabla~\cdot~\bm v=0$.  This property allows an alternative  representation of the superfluid velocity field
\begin{equation}\label{nablachi}
\bm v =(\hbar/M)\, \hat{\bm n} \times \bm \nabla \chi,
\end{equation}
where $\chi(\bm r)$ is the stream function and $\hat{\bm n}$ is the unit normal vector to the plane.  For a single quantized  vortex with integer vortex charge $q_0$ at $\bm r_0$, the vorticity is singular, with 
\begin{equation}\label{vorticity}
\bm \nabla\times \bm v = \frac{2\pi\hbar}{M} \,\hat{\bm n}\, q_0\delta^{(2)}(\bm r-\bm r_0).
\end{equation}
 It is not hard to show  from Eqs.~(\ref{nablachi}) and (\ref{vorticity}) that the scalar  function $\chi$ satisfies  Poisson's equation 
\begin{equation}\label{Poisson}
\nabla^2 \chi = 2\pi \sum_{j=1}^{N_v} \,q_j \,\delta^{(2)}(\bm r-\bm r_j),
\end{equation}
with the $N_v$ vortices at $\bm r_j$ and charges $q_j$  serving as sources for the stream function.  Note the clear analogy to the two-dimensional electrostatic potential arising from a set of two-dimensional point charges.   Like the electrostatic potential, the total stream function $\chi$  is defined only up to an additive constant that is generally irrelevant.

Equations (\ref{nablaS}) and (\ref{nablachi}) together give the following equations for the Cartesian components of the velocity field 
\begin{equation}\label{vxvy}
\frac{Mv_x}{\hbar} =  \frac{\partial S}{\partial x} = -\frac{\partial \chi}{\partial y};\quad\frac{Mv_y}{\hbar} =  \frac{\partial S}{\partial y} = \frac{\partial \chi}{\partial x}. 
\end{equation}
By inspection, we see that $\chi$ and $S$ satisfy the Cauchy-Riemann equations and can be linked to form a single analytic  function of a complex variable
\begin{equation}\label{F}
F(z) = \chi + i S,
\end{equation}
where $z = x+iy$;  its real part immediately provides the desired stream function with $\chi = {\rm Re} \,F $.  It also follows directly that 
\begin{equation}\label{Fprime}
v_y + i v_x = \frac{\hbar}{M}\frac{dF}{dz} = \frac{\hbar}{M}F'(z).
\end{equation}

The  complex potential for a single vortex with positive charge $q_0 = 1$ at $z_0$  is 
\begin{equation}\label{Fvortex}
F(z) =  \ln(z-z_0). 
\end{equation}
Any complex function $Z(z)$, such as  $z-z_0$, can be written in polar form: $Z(z) = |Z(z)| \exp[i\,{\rm arg}\, Z(z)]$, and note that  $|Z|= (ZZ^*)^{1/2}$, where $Z^*$ is the complex conjugate.  Correspondingly, the  logarithm becomes $\ln Z = \ln |Z| + i\, {\rm arg} Z$ and the real part is $\ln|Z| =  \frac{1}{2} \ln |Z|^2$.

  For a single positive vortex   [compare Eq.~(\ref{Fvortex})], the  stream function $\chi_{0}(\bm r) =  \,{\rm Re}\,F$ is 
\begin{equation}\label{chivortex}
\chi_{0}(\bm r) = {\textstyle\frac{1}{2}}\ln |z-z_0|^2 =  \ln|\bm r-\bm r_0|.
\end{equation}
The resulting superfluid velocity $\bm v(\bm r)$ follows from Eq.~(\ref{Fprime}) 
\begin{equation}
v_y + i v_x = \frac{\hbar}{M} \frac{1}{z-z_0} = \frac{\hbar}{M} \frac{z^*-z_0^*}{|z-z_0|^2}.
\end{equation} 
If the  vortex is at the origin ($z_0=0$), we readily recover the familiar axisymmetric circulating irrotational  flow 
\begin{equation}
\bm v_0(\bm r) = \frac{\hbar}{Mr}\,\hat{\bm \phi},
\end{equation}
 where $(r,\phi)$ are plane polar coordinates and $\hat{\bm \phi} = \hat{\bm n}\times \hat{\bm r}$ is the unit azimuthal vector in the polar direction.
 
 More generally, for a system of $N_v$ vortices at $\bm r_j$ with vortex charge $q_j$   ($j = 1,\,\cdots,N_v$), the total hydrodynamic flow field $\bm v(\bm r)$ is the sum of contributions from each vortex.  To make this connection more precise, define 
 \begin{equation}\label{chij}
 \chi_j(\bm r) =  \ln|z-z_j| =\ln|\bm r-\bm r_j|.
 \end{equation}
  We here consider only singly charged vortices with $q_j = \pm 1$, although most of our analysis applies more generally. The total stream function $\chi(\bm r)$  is the linear combination
 \begin{equation}\label{totalchi}
\chi(\bm r) = \sum_{j=1}^{N_v}q_j \chi_j(\bm r)
\end{equation}
giving  the total hydrodynamic flow field  
\begin{equation} \label{totalflow}
\bm v(\bm r) = \frac{\hbar}{M}\, \hat{\bm n}\times \bm \nabla \chi(\bm r) =   \frac{\hbar}{M}\, \hat{\bm n}\times \bm \nabla\sum_{j}q_j  \chi_j(\bm r).
\end{equation}

In an ideal fluid, a given vortex moves with the local flow velocity at its position, typically arising from all the {\em other} vortices.  In the present context of an unbounded plane, it follows directly that  the $k$th vortex has the velocity 
\begin{equation}\label{dynamics}
\dot{\bm r}_k = \frac{\hbar}{M}\, \hat{\bm n} \times {\sum_j}^{'}q_j  \left[\bm\nabla\chi_j(\bm r)\right]_{\bm r\to\bm r_k},
\end{equation}
where the primed sum omits the single term $j = k$.  For each term of the sum, the $j$th coordinate is fixed, and we can simplify this result to 
\begin{equation}\label{dynamicsplane}
\dot{\bm r}_k = \frac{\hbar}{M}\, \hat{\bm n} \times \bm\nabla_k { \sum_j}^{'}q_j \chi_j(\bm r_k),
\end{equation}
which now depends only on the coordinates of the vortices since $\chi_j(\bm r_k) = \ln|\bm r_k-\bm r_j|$.  In this way, the individual stream functions $\chi_j$  determine both the total hydrodynamic superfluid flow through Eq.~(\ref{totalflow}) and the dynamical motion of each vortex through Eq.~(\ref{dynamicsplane}).

The quantity $\bm\nabla_k \chi_j(\bm r_k)=(\bm r_k-\bm r_j)/|\bm r_k-\bm r_j|^2$ is odd under the interchange $j\leftrightarrow k$.  It follows immediately from Eq.~(\ref{dynamicsplane}) that 
\begin{equation}\label{conservation}
\sum_{k=1}^{N_v} q_k \dot{\bm r}_k = 0, \ \ \hbox{or, \ equivalently,}\ \  \sum_{k=1}^{N_v} q_k \bm r_k = {\rm const.}
\end{equation}
 Hence the combined dynamical motion of the vortices on a plane  conserves this vector quantity $\sum_{k=1}^{N_v} q_k \bm r_k$. This result is well known for two vortices, but its generalization will be useful in Sec.~VI.

\section{Stereographic projection for vortices on a sphere}

Most of the previous summary about vortices on a plane applies also to vortices on the surface of a sphere, but there is one important difference.  On a plane, a simple closed curve $\cal C$ divides the plane into an interior and an exterior, with a clear distinction between them.  For example, a point in the interior can serve as a local origin that defines a positive sense of rotation for $\cal C$.  The phase winding around $\cal C$ is the net vortex charge enclosed by $\cal C$, and the resulting lines of constant phase crossing $\cal C$ can extend unimpeded through the exterior region.

The situation is very different on the surface of a sphere because of the compact topology.  In particular, a closed curve $\cal C$ divides the spherical surface into two regions, but the choice of interior region appears arbitrary.  Choose a point in one of the two regions and use it to define the positive sense for $\cal C$, with a resulting phase winding number.  From the perspective of the other region, however, $\cal C$ has a negative sense with the negative of the same phase winding number.  This symmetry means that any collection of point vortices on the surface of a sphere  must have zero total vortex charge.   Sections~80 and 160 of \cite{Lamb} quote  this property in treating a vortex dipole on a sphere as the simplest case with zero net vortex charge, using without proof the method of stereographic projection to give the translational velocity of such a vortex dipole.  Here we provide a more complete derivation.

Consider
 a point vortex in a superfluid film  on the surface of a sphere of radius $R$. To determine the velocity field on the sphere, we use a stereographic projection~\cite{2} onto a complex plane. Familiar methods from Sec.~II provide the velocity field on the complex  plane, and the appropriate  coordinate transformation then determines the corresponding solution for the sphere.

 In the stereographic projection, we consider a tangent plane at the north pole.  Each point on  the sphere has a one-to-one  correspondence with a point on  tangent  plane, except for the point at the south pole. As seen from Fig.~1, a point on the sphere with spherical polar coordinates $(\theta,\phi)$ has the corresponding  complex coordinate on the plane 
 \begin{equation}\label{stereo}
 z=\rho e^{i\phi}\quad \hbox{with} \quad \rho=2R\tan{(\theta/2)}
 \end{equation}
   and the same azimuthal angle $\phi$. 

\begin{figure}[!htb]
\centering
\includegraphics[width=1.0\columnwidth]{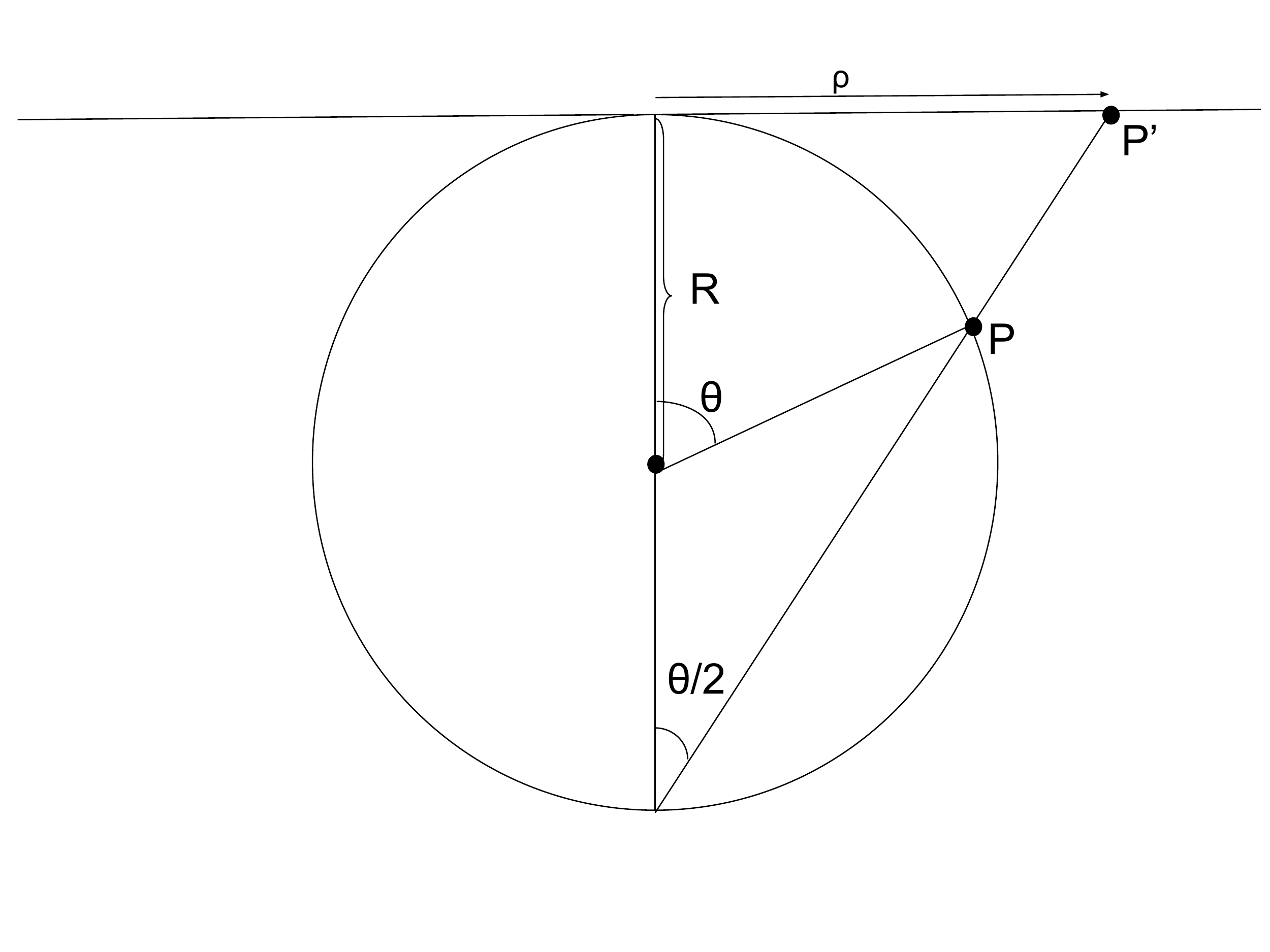}
\caption{Stereographic projection of a sphere of radius $R$ onto a tangent plane at the north pole. The figure shows the projection of a generic point $P$ on the sphere to the point $P'$ on the complex plane, with radial distance $\rho = 2R\tan(\theta/2)$ and the same azimuthal angle $\phi$.}
\label{Projecao estereografica}
\end{figure}

Section~II shows that the complex potential $F(z)$ of a vortex dipole with complex coordinates $z_\pm$ and charges $q_\pm = \pm 1$ is 
\begin{equation}\label{planedip}
F_{\rm dip}(z) =\ln(z-z_+)-\ln(z-z_-) = \ln\left(\frac{z-z_+}{z-z_-}\right),
\end{equation}
where $z$ is on the tangent plane.  The stereographic projection in Eq.~(\ref{stereo}) gives the transformation $z_\pm = 2R\tan(\theta_\pm/2)e^{i\phi_\pm}$, where $(\theta_\pm,\phi_\pm)$ are the spherical coordinates of the members of the vortex dipole on the sphere, leading to the associated complex function on the sphere
\begin{equation}
F_{\rm dip} = \ln\left(\frac{\tan(\theta/2)e^{i\phi} -\tan(\theta_+/2)e^{i\phi_+}}{\tan(\theta/2)e^{i\phi} -\tan(\theta_-/2)e^{i\phi_-}}\right),
\end{equation}
where the overall scale factors $2R$ cancel.

It is helpful to introduce the abbreviation $u = \tan(\theta/2)$ and similarly for $u_\pm$.  Application of Eq.~(\ref{chivortex}) for each vortex yields the stream function for a vortex dipole on a sphere 
\begin{equation}
\chi_{\rm dip}(\Omega) = \frac{1}{2} \ln\left(\frac{u^2 + u_+^2 - 2u u_+\cos(\phi - \phi_+)}{u^2 + u_-^2 - 2u u_-\cos(\phi - \phi_-)}\right),
\end{equation}
 where $\Omega  = (\theta,\phi)$  is a convenient notation. 
Use of the trigonometric identity 
\begin{equation}\label{chidip}
\tan(\theta/2) =\frac{\sin\theta}{1+\cos\theta} = \left(\frac{1-\cos\theta}{1+\cos\theta}\right)^{1/2}
\end{equation}
simplifies this result, which can be written  as  the difference of two separate quantities, each involving a single vortex [compare Eq.~(\ref{totalchi})]
\begin{equation}
\chi_{\rm dip}(\Omega )= \chi_+(\Omega) -\chi_-(\Omega).
\end{equation} 
 Here [compare Eq.~(\ref{chij})] 
\begin{eqnarray}\label{chijsphere}
\chi_j(\Omega) &=&  {\textstyle\frac{1}{2}}  \ln\left[2-2\cos\theta\cos\theta_j\right.\nonumber\\[.2cm]
& & \left. -2\sin\theta\sin\theta_j\cos(\phi-\phi_j)\right]
\end{eqnarray}
depends on the variable angular coordinate $\Omega$ and the fixed angular coordinate $\Omega_j$ of the $j$th vortex. 
Note that the factor $2$  merely adds a constant term to $\chi_j$ and  is chosen for convenience.  Also,  we omit a constant factor $1+\cos\theta_j$ that depends only on the vortex coordinate and does not affect the fluid velocity. Since Eq.~(\ref{chijsphere}) is symmetric in its variables, we can also write it as $\chi_j(\Omega) =  \chi(\Omega,\Omega_j)$, which will be useful below.

To interpret Eq.~(\ref{chijsphere}), recall  the unit radial vector $\hat{\bm r} = \sin\theta\cos\phi\,\hat{\bm x} +  \sin\theta\sin\phi\,\hat{\bm y} + \cos\theta\, \hat{\bm z}$. The dot product with the corresponding $\hat{\bm r}_j$ for the $j$th vortex is 
\begin{equation}
\hat{\bm r}\cdot\hat{\bm r}_j =\cos\gamma_j = \sin\theta \sin\theta_j\cos(\phi-\phi_j) + \cos\theta\cos\theta_j,
\end{equation}
where $\gamma_j$ is the angle between the two vectors.  Evidently, Eq.~(\ref{chijsphere}) has the equivalent and simpler form 
\begin{eqnarray}\label{chigammaj}
\chi_j(\Omega) &=& \chi(\Omega,\Omega_j) = {\textstyle\frac{1}{2}} \ln (2-2\cos\gamma_j) = \ln|\hat{\bm r}-\hat{\bm r}_j|\nonumber \\[.1cm] 
& =& {\textstyle\frac{1}{2}}  \ln\left[4\sin^2(\gamma_j/2)\right]=\ln\left[2\sin(|\gamma_j|/2)\right],
\end{eqnarray}
where the square root requires the absolute value $|\gamma_j|$ [compare Eq.~(\ref{chij}) for the stream function of a vortex on a plane].  

To provide a geometric interpretation, Fig.~1 shows that $2R\sin(\theta/2)$ is the chordal distance between the point $P$ and the north pole.  Hence $2R\sin(|\gamma_j|/2)$ is the chordal distance between $\hat{\bm r}$ and $\hat{\bm r}_j$, which is less than the corresponding great-circle distance between them.

For a general set of $N_v$ point vortices with $\sum_jq_j = 0$, the corresponding total stream function is
 \begin{equation}
\chi(\Omega) = \sum_j q_j\chi(\Omega,\Omega_j),
\end{equation}
which depends on the variable observation location $\Omega$ and the given locations of all the vortices $\Omega_j$.  The total hydrodynamic velocity field is 
\begin{equation}\label{vsphere}
\bm v(\Omega) = \frac{\hbar}{M}\, \hat{\bm r} \times \bm \nabla \chi(\Omega) =  \frac{\hbar}{M}\, \hat{\bm r} \times \sum_j q_j\bm \nabla\chi(\Omega,\Omega_j),
\end{equation}
where  $\hat{\bm r}$ is the unit normal vector to the surface of the sphere.  Note that the operator $\hat{\bm r} \times \bm \nabla$ is effectively the angular momentum operator and involves only the angular parts of the gradient operator.  To be very specific, 
\begin{equation}
\hat{\bm r}\times\bm\nabla = -\frac{\hat{\bm \theta}}{R\sin\theta} \frac{\partial}{\partial \phi } +  \frac{\hat{\bm\phi}}{R}\frac{\partial }{\partial\theta},
\end{equation}
where ($\hat{\bm \theta},\hat{\bm \phi},\hat{\bm r}$) form an orthonormal triad on the surface of the sphere, analogous to ($\hat{\bm x},\hat{\bm y},\hat{\bm n}$) on the plane.

Just as for a plane, a given vortex on a spherical surface moves with the local velocity generated by all the other vortices [see Eq.~(\ref{dynamics})]. 
The  argument leading to Eq.~(\ref{dynamicsplane})  readily yields
 \begin{equation}\label{dynamicssphere}
\dot{\bm r}_k = \frac{\hbar}{M}\, \hat{\bm r}_k \times\bm\nabla_k  {\sum_{j}}^{'}q_j  \chi(\Omega_k,\Omega_j).
\end{equation}
The dynamics of the $k$th vortex now depends only on the angular coordinates of the $N_v$ vortices through the single sum of stream functions 
\begin{equation}\label{sum}
{\sum_{j}}^{'} q_j\chi(\Omega_k,\Omega_j)={ \sum_{j}}^{'} q_j \ln\left[2\sin(|\gamma_{kj}|/2)\right],
\end{equation}
where $\cos\gamma_{kj} =\hat{\bm r}_k\cdot\hat{\bm r}_j$.

A single vortex dipole provides a simple example of vortex dynamics on a sphere.  
The rotational symmetry allows us to place the two vortices at the same azimuth angle $\phi_+=\phi_-$, with $\theta_+ $ in the upper hemisphere and $\theta_-$ in the lower hemisphere (we eventually choose them to be symmetrical around the equator, but we  first need to differentiate the stream function with respect to one polar angle, keeping the other polar  angle fixed).
Equation~(\ref{chigammaj}) gives the relevant stream function   
\begin{equation}\label{chidipsphere}
\chi(\Omega_+,\Omega_-) = \ln\left[2\sin\left(\frac{\theta_- -\theta_+}{2}\right)\right].
\end{equation}
Use of  Eq.~(\ref{dynamicssphere}) gives (compare Sec.~160 of \cite{Lamb})
 \begin{equation}\label{dynamicsdipole}
\dot{\bm r}_+ = \dot{\bm r}_- = \frac{\hbar}{2MR}\cot\left(\frac{\theta_- - \theta_+}{2}\right)\hat{\bm \phi} =  \frac{\hbar}{2MR}\tan\theta\,\hat{\bm \phi},
\end{equation}
where we set $\theta_+ = \theta$ and $\theta_-=\pi-\theta$ after applying the gradient operator.  Each member of the vortex dipole moves uniformly in the positive $\hat{\bm \phi}$ direction with fixed separation and speed $v_{\rm dip} = \hbar\tan\theta/(2MR)$.  For small $\theta\ll 1$, the dipole is near the poles and the motion is slow.  If the pair is near the equator, however, we have $\theta = \pi/2-\Delta\theta$ with $\Delta\theta\ll 1$, and $v_{\rm dip} = \hbar\cot\Delta\theta/(2MR)\approx \hbar/(2MR\Delta\theta)$.  This result agrees  with the corresponding planar dipole because $2R\Delta\theta$ is the linear separation between the two members of the dipole.

\section{Energy of vortices on a sphere}

In the present hydrodynamic  model, the total energy is simply the kinetic energy $E =\frac{1}{2}Mn\int d^2r |\bm v(\bm r)|^2$, where $n$ is the two-dimensional uniform number density  and the logarithmic divergence near the center of each vortex must be cut off at some small vortex core radius $\xi$. This picture applies equally to a plane and to the surface of a sphere.  A combination with Eq.~(\ref{vsphere}) leads to 
\begin{equation}
E = \frac{\hbar^2 n}{2M}\sum_{i,j=1}^{N_v}q_iq_j\int d^2 r\,\bm \nabla \chi_i\cdot\bm\nabla \chi_j  = \frac{\hbar^2 n}{2M}\sum_{i,j=1}^{N_v}q_iq_j \,I_{ij},
\end{equation}
which defines  the  dimensionless integrals $I_{ij} =\int d^2 r\,\bm \nabla \chi_i\cdot\bm\nabla \chi_j$.  The sum  includes both the diagonal terms with $i=j$ and the off-diagonal terms with $i\neq j$.

It is convenient to use the vector  identity 
\begin{equation*}
\bm\nabla A\cdot \bm\nabla B = \bm\nabla\cdot (A\bm \nabla B)-A\nabla^2 B,
\end{equation*} 
where $A$ and $B$ are scalar functions.  For the diagonal terms, we exclude a small circle of radius $\xi$ around $\bm r_j$, where the relative angle $\gamma_j$ is small and constant.  In this vicinity, Eq.~(\ref{chigammaj}) shows that $\chi_j \approx \frac{1}{2} \ln (\gamma_j^2)$, and a detailed analysis gives $I_{ii} = 2\pi \ln( R/\xi)$, where $R$ is the radius of the sphere.  Note that the resulting self-energy $E_{\rm self} = (\pi\hbar^2n/M) \ln(R/\xi)$ is independent of position.

For the off-diagonal integrals with $i\neq j$, the above vector identity and Eq.~(\ref{Poisson}) immediately gives the result $I_{ij} = -2\pi \chi_{ij}$, where we use the simplified notation $\chi_{ij} =  \chi(\Omega_i,\Omega_j)$.  We can now combine our results to find the total energy 
\begin{equation}\label{totalenergy}
E = \frac{\hbar^2n\pi }{M} N_v \ln\left(\frac{ R}{\xi}\right) -\frac{\hbar^2n\pi}{M}{\sum_{i,j}}^{'} q_iq_j\,\chi_{ij},
\end{equation}
where the primed double sum omits terms with $i = j$.  The first term (the total self-energy of the $N_v$ individual vortices) is an  irrelevant additive constant, whereas
 the second term (the interaction energy) depends explicitly on the position and sign of all the vortices because $\chi_{ij}= \frac{1}{2} \ln\left[4\sin^2(\gamma_{ij}/2)\right]$ involves the relative angle $\gamma_{ij}$ between the two vortices.

A combination with Eq.~(\ref{dynamicssphere}) leads to the important result 
\begin{eqnarray}\label{Edynamics}
2\pi \hbar n  q_k\dot{\bm r}_k& =&  \frac{2\pi \hbar^2n}{M}\hat{\bm r}_k\times\bm\nabla_k {\sum_j}^{'}q_kq_j\chi(\Omega_k,\Omega_j) \nonumber\\[.2cm]
&=&-\hat{\bm r}_k\times\bm\nabla_k E,
\end{eqnarray}
where we note that 
\begin{equation*}
 \bm\nabla_k {\sum_{i,j}}^{'} q_iq_j\chi(\Omega_i,\Omega_j) = 2\bm\nabla_k {\sum_j}^{'}q_kq_j\chi(\Omega_k,\Omega_j).
 \end{equation*}
   Also, the total self-energy is constant and hence does not contribute to the induced velocity.  
The quantity $-\bm\nabla_k E$ can be considered a force on the $k$th vortex.  The hydrodynamics of ideal fluids  then ensures that the vortex moves perpendicular to this  force, in contrast to the usual Newtonian situation for point masses.

Although Eq.~(\ref{Edynamics}) is correct and instructive as written, it can be helpful to separate it into spherical polar components on the sphere's surface.  
The velocity vector on the surface has the corresponding components 
\begin{equation}
\dot{\bm r} = R\dot\theta \,\hat{\bm \theta} + R\sin\theta \dot\phi\,\hat{\bm \phi}.
\end{equation}
Also, 
\begin{equation}
\bm \nabla_k E =  \frac{1}{R}\frac{\partial E}{\partial \theta_k}\hat{\bm \theta}_k  +  \frac{1}{R\sin\theta_k}\frac{\partial E}{\partial \phi_k}\hat{\bm \phi}_k 
\end{equation}
A straightforward comparison of terms shows that Eq.~(\ref{Edynamics}) has the equivalent form 
\begin{eqnarray}
2\pi\hbar n  q_k R \,\dot\theta_k & = &\frac{1}{R\sin\theta_k} \frac{\partial E}{\partial \phi_k}, \\
2\pi\hbar n q_k R \sin\theta_k\,\dot\phi_k & = &-\frac{1}{R} \frac{\partial E}{\partial \theta_k}.
\end{eqnarray}
As noted by Kirchhoff for point vortices on a plane (see \cite{Lamb}, Sec.~157), these equations have a Hamiltonian structure with $N_v$ pairs of  conjugate canonical variables $(\theta_k,\phi_k)$. For a sphere, $N_v$ must be even with $\sum_kq_k =0$.  In the present case, the polar motion $\dot\theta_k$ depends on $\partial  E/\partial \phi_k$, and similarly  the azimuthal motion $\dot\phi_k$ depends on $-\partial E/\partial \theta_k$, as seen for a single vortex dipole in Eq.~(\ref{dynamicsdipole}).

\section{Dynamics of four vortices arranged symmetrically}

In  Sec.~III, we determined the velocity field of a single vortex dipole on a sphere.  Equation~(\ref{dynamicsdipole}) shows that the dipole moves at constant  angular separation parallel to the great circle bisecting them in the direction of the fluid flow between them. The speed of the dipole depends only on the cotangent of the half the angular aperture between them. 

\subsection{Some simple configurations of four vortices}
 
The next simplest case is four vortices with overall charge neutrality.  Among many initial configurations, some have particularly simple dynamical trajectories.  Here we explore two coplanar initial configurations (symmetric and antisymmetric or exchanged) along with the three-dimensional tetrahedral configuration.  

\subsubsection{symmetric configuration $++--$}

\begin{figure}[!htb]
\centering
\includegraphics[width=0.8\columnwidth]{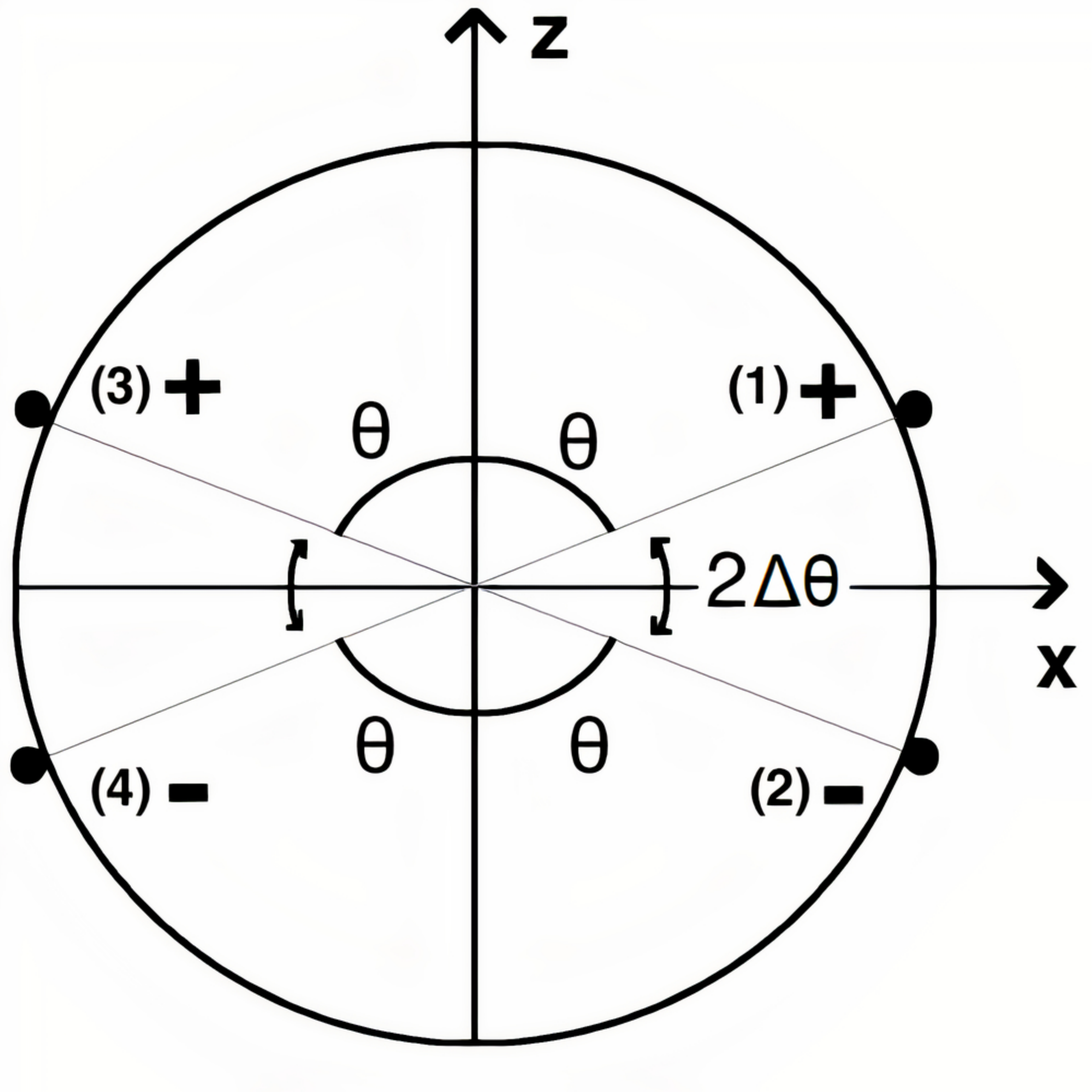}
\caption{Coplanar symmetric   configuration for four  vortices numbered  1 to 4 and  charged as shown.  We choose initial  azimuthal angles $\phi=0$ (vortices 1 and 2) and $\phi=\pi$ (vortices 3 and 4) and initial polar angles $\theta$ (vortices 1 and 3) and $\pi-\theta$ (vortices 2 and 4) measured from the $z$ axis. The figure shows the angular aperture $ 2 \Delta\theta = \pi - 2\theta$ for vortex dipoles (12) and (34), where $\Delta\theta = \pi/2-\theta$.}
\label{coplanar}
\end{figure}

Figure~\ref{coplanar} illustrates what we call the symmetric coplanar vortex configuration,  where we  choose the coordinate axes with all  four vortices initially  in the $xz$ plane. Vortices 1 (positive) and 2 (negative) 
 have  the same azimuthal coordinates $\phi_{1,2}=0$, and  symmetric polar coordinates $\theta_{1}=\theta$ and $\theta_{2}=\pi-\theta$ around the equator. The other 
 two vortices 3 (positive) and 4 (negative) are  mirror images of vortices 1 and 2, respectively, reflected in  the $yz$ plane, with  polar angles $\theta_3 = \theta$ and $\theta_4 = \pi - \theta$, and azimuthal coordinates $\phi_{3,4} = \pi$. 
 \\

These four vortices behave intuitively  in two simple limits:  When the vortices are near the poles ($\theta\ll 1$), the two positive (negative) vortices circle the north (south) pole, acting like independent pairs of same-sign vortices, with a linear speed $\hbar/(2MR\theta)$, just like two same-sign vortices on a plane.    Near the equator ($\theta\to \pi/2$ with $\Delta \theta=\pi/2-\theta  \ll 1$)  vortices ($1,2$) and ($3,4$)  act like two independent vortex dipoles and move uniformly with speed $\hbar/(2MR\Delta\theta)$,  straddling the equator.  For intermediate configurations, all four vortices influence each other, requiring a more detailed analysis.

Use of Eq.~(\ref{dynamicssphere}) shows that
\begin{eqnarray}
\label{vk} \nonumber
\dot{\bm r}_k &=& \sum _{ j\neq k}\frac{\hbar}{4 M R}\frac{q_j}{\sin^2(\gamma_{kj}/2)}\bigg\{-\left[\sin{\theta_j} \sin{(\phi_k-\phi_j)} \right] \hat{\bm\theta}_k \\   &+&\left[ \sin\theta_k\cos\theta_j- \cos\theta_k\sin\theta_j\cos(\phi_k-\phi_j) \right]\hat{\bm\phi}_k \bigg\}. 
\end{eqnarray}
For each $k = 1,\cdots,4$, a detailed analysis yields the simple result  
\begin{equation}\label{parallel}
\dot{\bm r}_k =  \frac{\hbar}{MR}\frac{1}{\sin{2\theta}}\, \hat{\bm\phi}_k  =  \frac{\hbar}{MR}\frac{1}{\sin{2\Delta\theta}}\, \hat{\bm\phi}_k,
\end{equation}
where $\Delta \theta = \pi/2 -\theta$.  Since these four translational velocities are equal and in the positive azimuthal direction, the vortices all rotate together in a positive (counterclockwise) sense around $\hat{\bm z}$, remaining coplanar.  In appropriate limits, Eq.~(\ref{parallel}) confirms our previous qualitative discussion.

\subsubsection{exchanged (antisymmetric) configuration $+-+-$}

\begin{figure}[!htb]
\centering
\includegraphics[width=0.8\columnwidth]{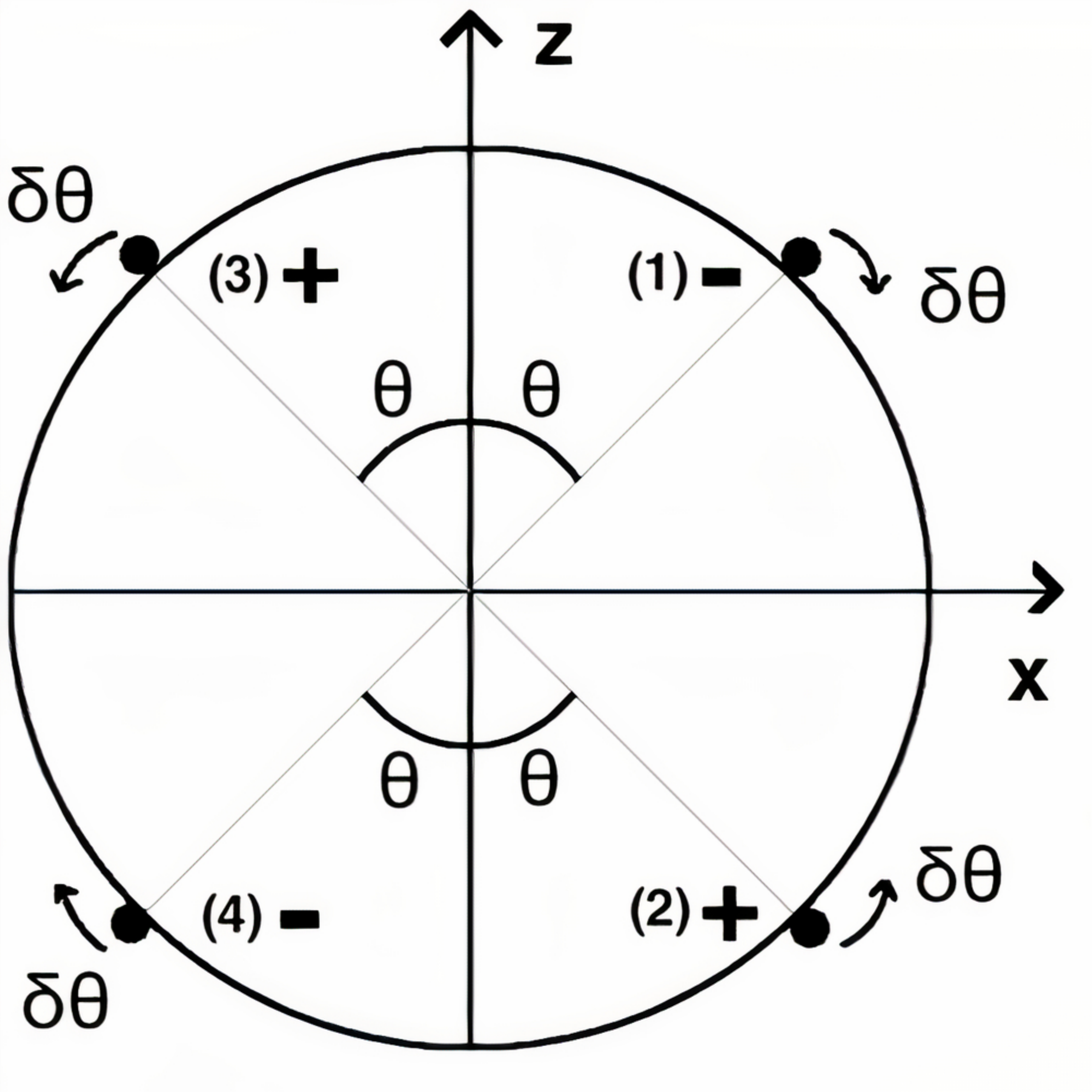}
\caption{Coplanar exchanged (antisymmetric) configuration for $\theta =  \pi/4$, including the small deviations $\delta\theta$ that provide the orbital motion around the equilibrium position. }
\label{exchange}
\end{figure}

  Consider now a second coplanar case, the exchanged configuration, where we interchange the charge of the vortices 1 and 2 in Fig.~\ref{coplanar}, with  $q_1 = -1$ and $q_2 = +1$, as shown in Fig.~\ref{exchange}.  
  
For simplicity, we start with $\theta \ll 1$, when the motion is intuitively clear.  Vortices (13) form a vortex dipole near the north pole, with vortices  (24) similarly forming a vortex dipole near the south pole.  Both dipoles move in the $yz$ 
plane at fixed $\phi = \pi/2$   toward the equator, where they interact and exchange partners.  Vortices (12) then move westward as a dipole along the equator, while vortices (34) similarly move eastward, meeting on the opposite side of the sphere. They then exchange partners again and move toward the north (south) poles, respectively, where the cycle repeats.  Note that the positive  (negative) vortices execute a close path in the positive (negative) sense relative to the outward normal to the surface of the  sphere (a video of the complete simulation is available in the Supplemental Material  \cite{SM} ).  Alternatively, we could have started with $\Delta\theta \ll 1$ (see Fig.~\ref{coplanar}), which executes the same cycle with a different initial position. 
As discussed  below Eq.~(\ref{dynamicsdipole}), each vortex moves with speed $\hbar/(2MR\theta)$ and travels a distance $2\pi R$ in one complete cycle.  Hence the period is $T= 4\pi MR^2\theta/\hbar$ and the corresponding angular frequency is 
\begin{equation}\label{omegasmall}
\omega = \frac{2\pi}{T}  = \frac{\hbar}{2MR^2\theta}.
\end{equation}

The other simple initial configuration is  $\theta = \pi/4$ (see Fig.~\ref{exchange}).  The sign exchange in the summed terms of Eq.~(\ref{vk}) means that  the velocities $\dot{\bm r}_k$ all vanish, yielding a static configuration of the four vortices. For small deviations of the initial positions with $\theta =  \pi/4 + \delta\theta$, numerical studies (see Supplemental Material  \cite{SM} )  show that  each vortex executes a closed elliptical orbit  around its equilibrium static  configuration, with  positive vortices moving in the positive sense and the negative ones in the negative sense, similar to the motion for the initial configuration $\theta \ll 1$.   
For $\delta\theta \ll  1$,  the elliptical orbit has semiminor and semimajor  axes   $\approx R \,2\delta\theta$ and $\approx R \,4\delta\theta$ along $\hat{\bm \theta}$ and $\hat{\bm \phi}$ directions, respectively. 
Using the numerical data for various values of $\delta\theta$, we found that the analytical formula 
\begin{equation}\label{omega}
\omega\approx \frac{\hbar}{M R^2}\frac{1}{\cos(2\delta\theta)}
\end{equation} 
provides a best fit to the numerical values of orbital frequency. 
For small $\delta \theta$, this frequency reduces to $\omega \approx \hbar/(MR^2)$, whereas for small $\theta\ll 1$ and $\delta\theta\to  -\pi/4$, Eq.~(\ref{omega}) gives $\omega \approx \hbar/(2 MR^2\theta)$, as expected for the orbital frequency of the dipoles in Eq.~(\ref{omegasmall}). The expression in Eq.~(\ref{omega}) may well be  exact, but we have not found a detailed proof.

For more general asymmetric $\theta$ configurations of the four vortices, they will rapidly  lose their initial coplanar condition, and we are unable to predict their orbits analytically.

Finally we  briefly discuss  a three-dimensional tetrahedral configuration, with the vortices at  four symmetric sites on the sphere.  In contrast to  four coplanar vortices, the tetrahedron has only one possible configuration:  each positive site will have one positive neighbor and two negative neighbors, all at the same separations.  Equation~(\ref{vk})  shows that the tetrahedron is also a static vortex configuration.  Unlike the planar configurations, however, it is an unstable equilibrium, with complicated dynamics even for small perturbations. 

\section{Energy and dynamics of two vortex dipoles}

Section IV considered the energy of a general neutral system of vortices on the surface of a sphere, as given in Eq.~(\ref{totalenergy}).  The next Sec.~V then studied the dynamics of four vortices with zero net vortex charge in some simple  symmetric configurations.  Here, we examine the case of  two small vortex dipoles, which is a different special configuration of four vortices with zero net vortex charge.   For definiteness, we consider one dipole with charges $q_1=-q_2 = 1$ and another dipole  with $q_3 = -q_4 =1$.  Although Eq.~(\ref{totalenergy}) gives the total interaction energy of these two dipoles as a sum over six pairs of vortices, we instead provide a more physical picture.


 As a first step, we focus on the  dipole with vortices at angular positions $\Omega_1$ and $\Omega_2$ on the sphere.  The total stream function is 
\begin{equation}\label{dipole12}
\chi^{12}(\Omega) = \chi(\Omega,\Omega_1) - \chi(\Omega,\Omega_2),
\end{equation}
where Eq.~(\ref{chijsphere}) gives the stream function for each individual vortex.
For the present case of a small vortex dipole centered at $\Omega_0$, we write $\delta\Omega_0 = (\delta\theta_0,\delta\phi_0) = \Omega_1-\Omega_2$ and expand Eq.~(\ref{dipole12}) to first order in the small angular separations.  Define the vortex  dipole moment (a vector)
\begin{equation}
\bm p_0 = R\delta\theta_0\,\hat{\bm \theta}_0 + R\sin\theta_0\,\delta\phi_0\,\hat{\bm \phi}_0,
\end{equation}
which has the dimension of a length, with  $\hat{\bm \theta}_0$ and $\hat{\bm \phi}_0$ the unit spherical-polar vectors at $\Omega_0$ on the surface of the sphere.
It is not difficult to find 
\begin{equation}\label{spheredipole}
\chi_{\rm dip}(\Omega) \approx  -\frac{\bm p_0 \cdot R(\hat{\bm r}-\hat{\bm r}_0)}{R^2(\hat{\bm r}-\hat{\bm r}_0)^2} =   -\frac{\bm p_0 \cdot ({\bm r}-{\bm r}_0)}{({\bm r}-{\bm r}_0)^2},
\end{equation}
where $\bm r = R\hat{\bm r}$ is the coordinate vector normal to the surface of the sphere and we note that $\bm p_0\cdot \hat{\bm r}_0 = 0$.  This expression is completely analogous to the electrostatic potential arising from an electric dipole moment $\bm p$, here suitably modified for the two-dimensional character of a vortex dipole.

The interaction energy of this single vortex dipole follows directly from Eq.~(\ref{totalenergy}) and the subsequent discussion
\begin{equation}
E_{12} = \frac{2\hbar^2n\pi}{M} \chi_{12}=\frac{2\hbar^2n\pi}{M} \ln\left[2\sin\left(\frac{|\gamma_{12}|}{2}\right)\right].
\end{equation}
For a small dipole with $|\gamma_{12}|\ll 1$, we find the simple expression
\begin{equation}\label{E12}
E_{12} \approx \frac{2\hbar^2n\pi}{M} \ln\left(|\gamma_{12}|\right) = -\frac{2\hbar^2n\pi}{M} \ln\left(\frac{R}{p_{12}}\right), 
\end{equation}
where $p_{12}= R|\gamma_{12}|$ is the separation between the two vortices (it is also the dipole moment).  As a result, the  energy of the two separate dipoles is
\begin{equation}\label{Ed}
E_{\rm d} = \frac{2\hbar^2 n\pi}{M}\left(\chi_{12} + \chi_{34}\right).
\end{equation}




The interaction energy of the two dipoles is $E_{\rm dd} = Mn\int d^2r \,\bm v^{12}\cdot\bm v^{34}$, where $\bm v^{12}$ and $\bm v^{34}$ are the velocity fields of the two dipoles.  Formally, we can use Eq.~(\ref{dipole12}) and write 
\begin{eqnarray}\label{Edd}
E_{\rm dd} &=&\frac{\hbar^2 n}{M}\int d^2 r \,\bm\nabla \chi^{34}\cdot \bm \nabla\chi^{12}\nonumber \\
&=& \frac{\hbar^2 n}{M}\int d^2 r \,\bm\nabla(\chi_3-\chi_4)\cdot\bm\nabla(\chi_1-\chi_2)\nonumber\\
  &=&-\frac{2\pi\hbar^2 n}{M}(\chi_{13} + \chi_{24} -\chi_{23}-\chi_{14}).
\end{eqnarray}
The sum of Eqs.~(\ref{Ed}) and (\ref{Edd}) precisely reproduces the six terms inferred from Eq.~(\ref{totalenergy}).

More physically, we can find $\chi^{12}$  and $\chi^{34}$ directly from Eq.~(\ref{spheredipole}).  In this way, the familiar  integration by parts with the first line of Eq.~(\ref{Edd})
 gives an explicit expression for the interaction energy of two small vortex dipoles $\bm p$ and $\bm p'$ at positions $\bm r$ and $\bm r'$:
 \begin{equation}\label{Epp}
E_{\rm dd} = \frac{2\pi \hbar^2n}{M} \,\frac{(\bm p\cdot\bm p')\, (\bm r-\bm r')^2 - 2\bm p\cdot(\bm r-\bm r')\,\bm p'\cdot(\bm r-\bm r')}{(\bm r-\bm r')^4}. 
\end{equation}
This rather complicated form is familiar from electrostatics and magnetostatics,  suitably modified for the two dimensions relevant here.  It varies inversely with the squared separation of the dipoles and is highly anisotropic through the dependence on each  dipole's orientation.  

We emphasize that Eq.~(\ref{Epp})  only applies for large separations $|\bm r-\bm r'|^2 \gg pp'$ and does not hold for two nearby dipoles.   In such a case, we must rely on the general expression from Eq.~(\ref{totalenergy}), involving a sum over all six distinct pairs of vortices.


It is interesting to consider the dynamics of two small vortex dipoles on a sphere, say $\bm p_i$ and $\bm p_i'$.  When  the two dipoles are well separated,  the total energy is the sum of their individual dipole energies $E_{\rm tot} = -(\pi\hbar^2n/M)\ln(R^2/p_ip_i')$, and the magnitudes $p_i$ and $p_i'$ remain fixed as each moves along a great circle at speed $\hbar/Mp_i$ and $\hbar/Mp_i'$, respectively

If they approach each other, the dynamics  becomes complicated, requiring a detailed analysis.  Nevertheless, some general conclusions are possible.  For two small nearby dipoles, they interact locally on the tangent plane.  We can then invoke the conservation law from Eq.~(\ref{conservation}), which now implies that the total dipole moment is conserved along with the total energy.  In an obvious notation for initial and final dipole moments, we have $\bm p_i+ \bm p_i' = \bm p_f + \bm p_f'$ and the conservation  of energy  implies that $p_ip_i' = p_fp_f'$.  

\begin{figure}[!h]
\centering
\includegraphics[width=1.0\columnwidth]{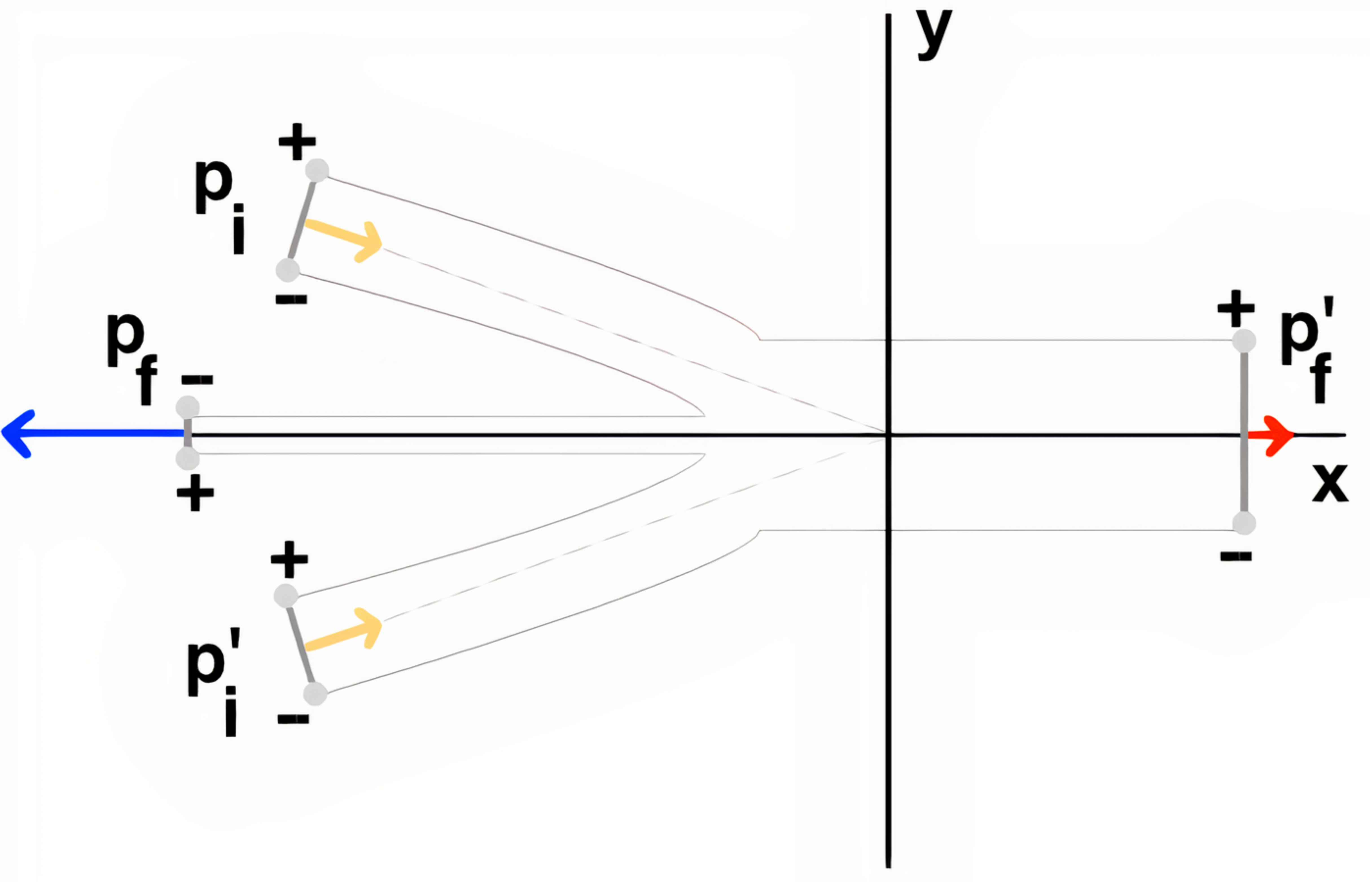}
\caption{Two vortex dipoles  $\bm p_i$ and $\bm p_i'$ move to the right on  nearly parallel converging  trajectories.  Once they are sufficiently close, the individual vortices  rearrange to form two new vortex dipoles, with $\bm p_f$ moving rapidly to the left and $\bm p_f'$ moving slowly to the right. The solid lines show calculated orbits for the initial conditions.}
\label{dipoles}
\end{figure}

We already discussed a specific example of two dipoles on a sphere  in connection with Fig.~3, when one dipole starts at the north pole and the other starts at the south pole as shown for small dipoles with $\theta \ll 1$.  Here we assume that $p_i'= p_i= 2R\theta$.  After the interaction, the reconstructed dipoles move around the equator in opposite directions, and the  conservation laws then show that $p_f =p_f' =2R\theta$  during this part of their orbits.

A more unusual situation arises for two nearly parallel dipoles that gradually approach each other (see Fig.~4) in the tangent plane, labeled by axes $x$ and $y$.  Here the initial dipole moments are $\bm p_i\approx \bm p_i' \approx p\hat{\bm y}$, and their motion is principally along $\hat{\bm x}$ (converging  slowly  toward the $x$ axis), with each having a small velocity component  along $\hat{\bm y}$.  After the interaction, the two inner vortices form a new smaller dipole 
$\bm p_f = -p_f \hat{\bm y}$ that moves to the left.  In contrast,  the two outer vortices  form a new larger dipole $\bm p_f' = p_f' \hat{\bm y}$ that continues moving to the right.  The conservation of dipole moment requires $2p = p_f'-p_f$, and the conservation of energy shows that $p^2 = p_fp_f'$.  Simple algebra yields the final values $p_f'/p  = \sqrt 2+1$ and $p_f/p =\sqrt 2 -1$.  As a result, we have a very  asymmetric configuration with $p_f'/p_f = 3+ 2\sqrt2\approx 5.83$.  These  two new dipoles move in opposite directions around a single great circle at different speeds, with $v_f/v_f' = 3+ 2\sqrt2\approx 5.83$, but they eventually meet.  After interacting a second time, they  separate and move  on nearby great circles toward their initial configuration (see Supplemental Material \cite{SM} for a video of the early evolution of two colliding dipoles).

\section{Summary and Conclusions}

 We presented the dynamics and stability of singly charged vortices in a thin spherical ``bubble" BEC. 
 The topology of this system constrains the configuration of vortices to an even number of vortices 
 with total charge zero.  We calculated the velocity fields and energies of various configurations of vortices and dipoles, which allow us to determine their dynamics on the surface of the sphere.  In some cases, two initial small vortex dipoles can interact and exchange partners. The final reconfigured dipoles then separate and follow new trajectories.

 Proposed new  experiments in  microgravity conditions \cite{David2020}  means that our model of a  thin ``bubble" trapped BEC  may be studied experimentally in the near future.  
   In addition, recent experiments have studied a ``bubble" trap potential  in the presence of gravity \cite{Rossi,Guo}, and it may  soon be feasible  to  manipulate the trap rotation and the atomic parameters to fix various  initial conditions. We anticipate that our theoretical studies will be relevant  for any further studies on the evolution of vortex pairs  on a sphere.
   We plan to provide theoretical explanations for these experimental results, adapting our model to include the effect of the gravitational potential~\cite{Sun}.   

\section{Acknowledgments} We are grateful to Pietro Massignan for helpful discussions. This work was supported by the S\~{a}o Paulo Research Foundation (FAPESP) under Grant No. 2013/07276-1. The authors also thank Coordena{\c c}\~{a}o de Aperfei{\c c}oamento de Pessoal de N\'{i}vel Superior (CAPES) for their financial support.

\bibliography{references}

\begin{widetext}

\section*{Supplemental Material: Superfluid vortex dynamics on a  spherical film}

See \url{https://youtu.be/QqWxwSvG2o8} for a video displaying a complete simulation of the dynamics
of two vortex-dipoles. We start with $\theta \ll 1$ (see Fig.~3 in the main text).  The vortex pair black (positively charged) - red (negatively charged) forms a vortex dipole near the north pole. Similarly, the blue (positive) - orange (negative) vortices form a vortex dipole near the south pole.  Both dipoles move in the $yz$  plane  toward the equator, where they interact and exchange partners.  The new dipoles then move along the equator, meeting on the opposite side of the sphere. They then exchange partners again, with one of the new dipoles moving toward the north pole and the other towards the south pole, restarting the cycle.  \\ 
Moreover, see \url{https://youtu.be/U2b2oKtFjMA} for a video displaying a simulation of the vortices' orbits for small deviations of their initial positions with $\theta =  \pi/4 + \delta\theta$ (see Fig.~3). Each vortex executes a closed elliptical orbit  around its equilibrium static  configuration, with  positive vortices (black and blue) moving in the positive sense and the negative ones (red and orange) in the negative sense, similar to the motion for the initial configuration $\theta \ll 1$.   

See \url{https://youtu.be/GP6VTKZOm7A} for a video of the collision of two asymmetric dipoles (different speeds) with parameters set as in Fig.~4 of the main text. Initially, the slow pair $``$black (positively charged) - red (negatively charged)$"$ form a vortex dipole near the north pole, with the fast pair  $``$blue (positive) - orange (negative)$"$ forming a vortex dipole near the south pole.
After the interaction, the two initial dipoles exchange their components and form two equal-size (same speed) dipoles which move slightly tipped great circles that intersect on the opposite side of the sphere. Then they swap pairs again and reconstruct the initial dipoles configuration. This process repeats cyclically.

The unit of length in these simulations is the radius of the sphere $R$ and the unit of time is $M R^2/\hbar$, with $M$ being the atomic mass. 

\end{widetext}

\end{document}